\documentclass{emulateapj}

\usepackage{amsmath}
\usepackage{lscape}
\usepackage{multirow}

\newcommand{\about}{$\sim\!\!$~}

\newcommand{\be}{\begin{displaymath}}
\newcommand{\ee}{\end{displaymath}}

\def\lsim{\hbox{\rlap{\raise 0.425ex\hbox{$<$}}\lower 0.65ex\hbox{$\sim$}}}
\def\gsim{\hbox{\rlap{\raise 0.425ex\hbox{$>$}}\lower 0.65ex\hbox{$\sim$}}}

\def\arcsec{\hbox{$^{\prime\prime}$}}
\def\arcdeg{\mbox{$^\circ$}}

\newcommand{\msun}{M$_\sun$}

\newcommand{\etal}{et al.\ }

\newcommand{\halpha}{H$\alpha$}

\newcommand{\hbeta}{H$\beta$}

\newcommand{\kms}{km~s$^{-1}$}

\graphicspath{{/Users/jsilv/ptf11kx/plots/}}

\shorttitle{Late-Time Observations of PTF11kx}
\shortauthors{Silverman, et al.}

\begin{document}

\title{Late-Time Spectral Observations of the Strongly Interacting
  Type Ia Supernova PTF11kx}

\author{Jeffrey M. Silverman\altaffilmark{1,2,3}, Peter
  E. Nugent\altaffilmark{2,4}, Avishay Gal-Yam\altaffilmark{5},
  Mark Sullivan\altaffilmark{6}, D. Andrew Howell\altaffilmark{7,8},
  Alexei V. Filippenko\altaffilmark{2}, Yen-Chen Pan\altaffilmark{9},
  S. Bradley Cenko\altaffilmark{2}, and Isobel
  M. Hook\altaffilmark{9,10}} 

\altaffiltext{1}{Department of Astronomy, University of Texas, Austin, TX 78712-0259, USA.}
\altaffiltext{2}{Department of Astronomy, University of California, Berkeley, CA 94720-3411, USA.}
\altaffiltext{3}{email: jsilverman@astro.as.utexas.edu.}
\altaffiltext{4}{Lawrence Berkeley National Laboratory, Berkeley, CA 94720, USA.}
\altaffiltext{5}{Benoziyo Center for Astrophysics, The Weizmann Institute of Science, Rehovot 76100, Israel.}
\altaffiltext{6}{School of Physics and Astronomy, University of Southampton, Southampton, SO17 1BJ, UK.}
\altaffiltext{7}{Las Cumbres Observatory Global Telescope Network, Goleta, California 93117, USA.}
\altaffiltext{8}{Department of Physics, University of California, Santa Barbara, CA 93106, USA.}
\altaffiltext{9}{Department of Physics (Astrophysics), University of Oxford, Keble Road, Oxford OX1 3RH, UK.}
\altaffiltext{10}{INAF, Osservatorio Astronomico di Roma, Via Frascati 33, 00040, Monte Porzio Catone (RM), Italy.}

\begin{abstract}
PTF11kx was a Type~Ia supernova (SN~Ia) that showed time-variable
absorption features, including saturated \ion{Ca}{2}~H\&K lines that
weakened and eventually went into emission. The strength of the
emission component of \halpha\ gradually increased, implying that the
SN was undergoing significant interaction with its circumstellar
medium (CSM). These features, and many others, were blueshifted
slightly and showed a 
P-Cygni profile, likely indicating that the CSM was directly related
to, and probably previously ejected by, the progenitor system
itself. These and other observations led Dilday et al.\ (2012) to
conclude that PTF11kx came from a symbiotic nova progenitor
like RS~Oph. In this work we extend the spectral coverage of PTF11kx
to 124--680 rest-frame days past maximum brightness. The late-time 
spectra of PTF11kx are dominated by \halpha\ emission (with widths 
of full width at half-maximum intensity $\approx 2000$~\kms), strong
\ion{Ca}{2} emission features 
(\about10,000~\kms\ wide), and a blue ``quasi-continuum'' due to many
overlapping narrow lines of \ion{Fe}{2}. Emission from oxygen,
\ion{He}{1}, and Balmer lines higher than \halpha\ is weak or
completely absent at all epochs, leading to large observed
\halpha/\hbeta\ intensity ratios. The \halpha\ emission appears to 
increase in strength with time for \about1~yr, but it subsequently
decreases significantly along with the \ion{Ca}{2} emission. Our
latest spectrum also indicates the possibility of newly formed dust in
the system as evidenced by a slight decrease in the red wing of
\halpha. During the same epochs, multiple narrow emission features
from the CSM temporally vary in strength. The weakening of
the \halpha\ and \ion{Ca}{2} emission at late times is possible
evidence that the SN ejecta have overtaken the majority of the CSM and
agrees with models of other strongly interacting SNe~Ia. The varying
narrow emission features, on the other hand, may indicate that the CSM
is clumpy or consists of multiple thin shells. 
\end{abstract}

\keywords{supernovae: general --- supernovae: individual (PTF11kx) ---
  stars: circumstellar matter}


\section{Introduction}\label{s:intro}
Type~Ia supernovae (SNe~Ia), the result of the thermonuclear explosion
of C/O white dwarfs (WDs), provided the first clear evidence of the
Universe's accelerating expansion \citep{Riess98:lambda,Perlmutter99}
and have been used to measure various cosmological parameters 
\citep[e.g.,][]{Hicken09:cosmo,Conley11,Sullivan11,Suzuki12}. The two
main progenitor scenarios that likely lead to SNe~Ia are the
single-degenerate (SD) channel, when the WD accretes matter from a
nondegenerate companion star \citep[e.g.,][]{Whelan73}, and the
double-degenerate (DD) channel, which is the result of the merger of
two WDs \citep[e.g.,][]{Iben84,Webbink84}.

While it is unclear how often either of these scenarios occur, it
seems likely that both are actually present in Nature. For a handful of
extremely nearby SNe~Ia, many plausible SD scenarios have been ruled
out
\citep[e.g.,][]{Nugent11,Foley12,Bloom12,Silverman12:12cg,Schaefer12},
and the so-called super-Chandrasekhar mass SNe~Ia likely result from
the DD scenario
\citep[e.g.,][]{Howell06,Yamanaka09,Scalzo10,Silverman11,Taubenberger11}. 
On the other hand, photoionization and subsequent recombination of the
circumstellar medium (CSM) created by the progenitor system itself has
been observed in a few relatively 
normal SNe~Ia \citep{Patat07,Blondin09,Simon09}, and CSM has likely
been detected in the spectra of at least 20\% of SNe~Ia in spiral
galaxies \citep{Sternberg11}. Furthermore, extreme interaction with
CSM has been observed in a relatively small number of overluminous
SNe~Ia mainly via the detection of strong \halpha\ emission. These
``hybrid'' objects also resemble Type~IIn SNe (SNe~IIn) and have been
dubbed SNe~Ia-CSM \citep{Silverman13:Ia-CSM}.

Previously, it was not completely clear whether these objects
are actually SNe~Ia or instead a new subtype of core-collapse SN
\citep[e.g.,][]{Benetti06,Trundle08}. This controversy seems to
have been settled by the discovery and analysis of PTF11kx
\citep{Dilday12}. Discovered on 16~Jan.~2011 (UT dates
are used throughout this paper) by the Palomar Transient Factory
\citep[PTF;][]{Rau09,Law09} at redshift $z=0.0466$
\citep{Dilday12} and with Galactic reddening $E(B-V)=0.052$~mag 
\citep{Schlegel98}, it was 
shown to initially resemble the  
somewhat overluminous Type~Ia SN~1999aa
\citep{Li01:pec,Strolger02,Garavini04}, 
though with saturated \ion{Ca}{2}~H\&K absorption lines and weak
\ion{Na}{1}~D lines. This implies the object was almost certainly
a SN~Ia that had significantly affected its immediate surroundings
beginning shortly after explosion.

Subsequent spectra of PTF11kx presented by \citet{Dilday12} showed
time-variable absorption features of \ion{Na}{1}, \ion{Fe}{2},
\ion{Ti}{2}, and \ion{He}{1}, which (except for \ion{Na}{1}) had not
been seen in any previously observed SN~Ia. In addition, PTF11kx
revealed a strong \halpha\ line with a P-Cygni profile (indicative of an 
expanding shell of material) whose emission component gradually increased 
in strength,
causing the spectra of PTF11kx to eventually resemble those of other
Ia-CSM objects \citep{Silverman13:Ia-CSM}. The observations of
\citet{Dilday12} indicate the presence of multiple CSM components with
slower-expanding material exterior to faster material and with
velocities of \about50--100~\kms. Recent high-resolution observations
of RS~Oph \citep{Patat11} and models of circumstellar shells created
in such systems \citep{Moore12} seem to match many of the PTF11kx
observations; thus, \citet{Dilday12} suggest that PTF11kx was a
{\it bona fide} SN~Ia with a symbiotic nova progenitor \citep[i.e., a SD
scenario, but see][]{Shen13}. Extending at least some of these
findings to {\it all} of the Ia-CSM objects such that we can say that
they are all SNe~Ia, likely coming from a SD system, perhaps in the
form of a symbiotic nova scenario, is the goal of \citet{Silverman13:Ia-CSM}.

In \S\ref{s:spectra} we present eight late-time spectra of PTF11kx,
starting with the last spectrum shown by \citet{Dilday12} from
124 rest-frame days past maximum brightness and continuing through 680~d 
past maximum. We measure various spectral features and 
analyze the spectra in \S\ref{s:analysis}, and in \S\ref{s:comparisons} 
we compare the spectra of PTF11kx with those of other SNe. We summarize 
our conclusions in \S\ref{s:conclusions}.


\section{Spectra}\label{s:spectra}

Figure~1 of \citet{Dilday12} shows spectra of PTF11kx from 3~d before
$B$-band maximum brightness (which was on 29~Jan.~2011) to 88~d past
maximum, though one extra spectrum (observed 130~d past maximum, 
corresponding to 124~d in the rest frame) is 
listed in their Table~S2. This observation is the first one presented
in this work, where we extend the spectral coverage of PTF11kx to
680 rest-frame days past maximum brightness.
 
\begin{figure*}
\centering
\includegraphics[width=6.8in]{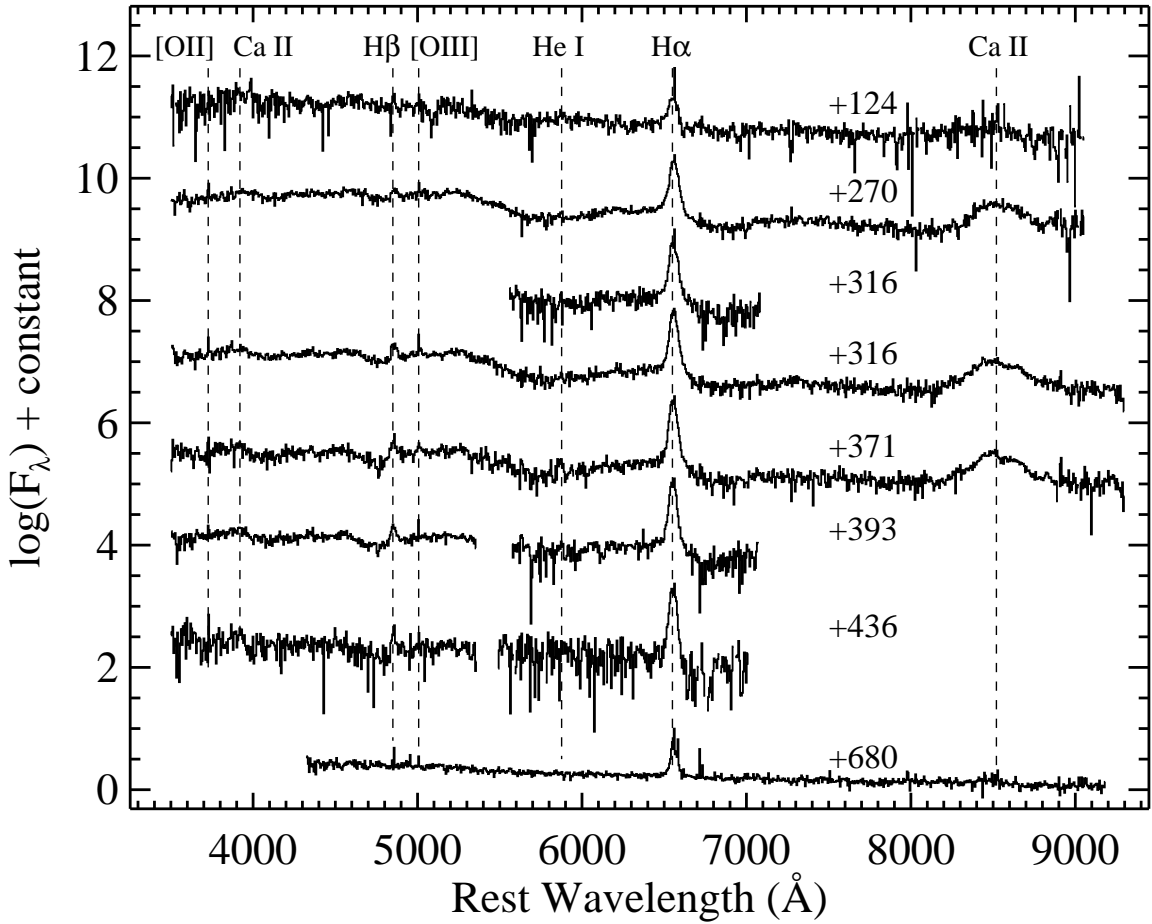}
\caption{Spectra of PTF11kx, labeled with rest-frame age relative to maximum
  brightness. The locations of detected spectral features, or possible
  spectral features, are labeled. The data have had
  their host-galaxy recession velocity 
  removed and have been corrected for Galactic reddening.}\label{f:ptf11kx}
\end{figure*}

Low-resolution optical spectra were obtained with the Intermediate 
dispersion Spectrograph and Imaging System
(ISIS)\footnote{http://www.ing.iac.es/Astronomy/instruments/isis/index.html .}
on the 4.2~m William Herschel Telescope (WHT), the Low Resolution Imaging
Spectrometer \citep[LRIS;][]{Oke95} on the Keck-I 10~m telescope, and
the DEep Imaging Multi-Object Spectrograph \citep[DEIMOS;][]{Faber03}
on the Keck-II 10~m telescope. All spectra were reduced using standard
techniques \citep[e.g.,][]{Silverman12:BSNIPI}. Routine CCD processing
and spectrum extraction were completed with IRAF\footnote{IRAF: The Image
  Reduction 
  and Analysis Facility is distributed by the National Optical Astronomy
  Observatory, which is operated by the Association of Universities
  for Research in Astronomy (AURA) under cooperative agreement with
  the National Science Foundation (NSF).}, and flux calibration
and removal of telluric lines were done using our own IDL routines.
Table~\ref{t:ptf11kx} lists information
regarding the PTF11kx spectra analyzed here, and the data are plotted
in Figure~\ref{f:ptf11kx}. Upon publication, all spectra presented in
this paper will be available in electronic format on WISeREP
\citep[the Weizmann Interactive Supernova data
REPository;][]{Yaron12}.\footnote{http://www.weizmann.ac.il/astrophysics/wiserep .} 

\begin{table*}
\begin{center}
\caption{Spectra of PTF11kx}\label{t:ptf11kx}
\begin{tabular}{lcllrr}
\hline
\hline
UT Date &  Age (d)$^\textrm{a}$  &  Instrument$^\textrm{b}$ & Range (\AA)  & Res. (\AA)$^\textrm{c}$ &  Exp. (s) \\
\hline
2011~Jun.~8 & 124 & ISIS & 3500--9498 & 3.5/7.2 & 3600 \\
2011~Nov.~7 & 270 & ISIS & 3500--9500 & 3.5/7.2 & 1800 \\
2011~Dec.~26 & 316 & LRIS & 5792--7440 & 3 & 600 \\
2011~Dec.~26 & 316 & LRIS & 3402--10148 & 3.9/6.1 & 900 \\
2012~Feb.~21 & 371 & LRIS & 3272--10250 & 3.8/6.0 & 655 \\
2012~Mar.~15 & 393 & LRIS & 3275--5628, 5809--7427 & 3.6/3 & 600 \\
2012~Apr.~29 & 436 & LRIS & 3236--5630, 5730--7360 & 3.7/3 & 625 \\
2013~Jan.~9 & 680 & DEIMOS & 4500--9636 & 3 & 1800 \\
\hline\hline
\multicolumn{6}{p{6.1in}}{$^\textrm{a}$Rest-frame days relative to
  $B$-band maximum brightness (29~Jan.~2011).} \\
\multicolumn{6}{p{6.1in}}{$^\textrm{b}$ISIS = Intermediate dispersion
  Spectrograph and Imaging System on the 4.2~m William Herschel
  Telescope; LRIS = Low Resolution Imaging Spectrometer on the Keck-I
  10~m telescope; DEIMOS
  = DEep Imaging Multi-Object Spectrograph on the Keck-II 10~m
  telescope.} \\
\multicolumn{6}{p{6.1in}}{$^\textrm{c}$Approximate resolution, full width 
  at half-maximum intensity (FWHM). If two numbers are listed,
  they represent the blue-side and red-side resolutions, respectively.} \\ 
\hline\hline
\end{tabular}
\end{center}
\end{table*}

Even though more than 550 rest-frame days are spanned by the data shown
in Figure~\ref{f:ptf11kx}, there is not much spectral evolution (until
the final spectrum presented). Two of our highest signal-to-noise
ratio (S/N) observations are separated by \about100~d and are nearly
identical (+270 and +371~d). Most of the emission features have
disappeared in the latest spectrum shown, from 680~d past maximum
brightness. All that is left is the relatively blue
``quasi-continuum,'' which is likely produced by emission from many
overlapping narrow lines of iron-group elements (IGEs, mostly
\ion{Fe}{2}) excited by the CSM interaction, and \halpha\ emission,
which has weakened since the previous spectrum (see below for 
further details). Conspicuously, the \ion{Ca}{2} near-infrared (IR) 
triplet, which is one of the strongest emission features in most of 
the late-time PTF11kx spectra, is almost completely gone by day +680.


\section{Analysis}\label{s:analysis}

\begin{figure*}
\centering
\includegraphics[width=6in]{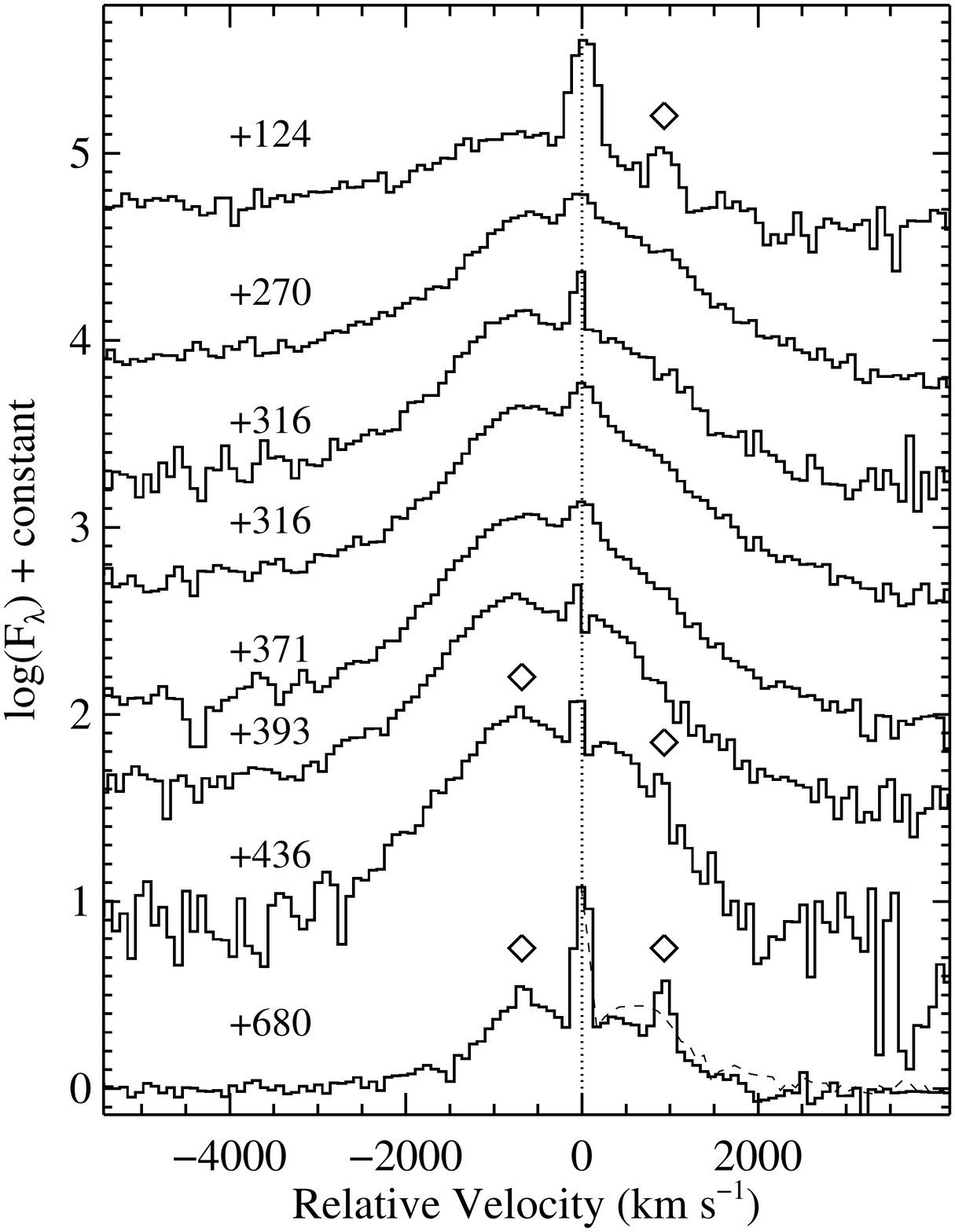}
\caption{The \halpha\ profiles of our spectra of PTF11kx,
  labeled with age relative to maximum 
  brightness. The data have had their host-galaxy recession velocity
  removed and have been corrected for Galactic reddening. The dotted
  vertical line is the systemic velocity of PTF11kx. The small
  emission features marked with diamonds are [\ion{N}{2}]
  $\lambda$6548.05 and $\lambda$6583.45 from the host galaxy. The
  dashed line on the bottom spectrum is the reflection of the blue
  half of the \halpha\ profile across the peak flux (after removing
  the [\ion{N}{2}] emission).}\label{f:hal_zoom} 
\end{figure*}

To better characterize the spectral features seen in PTF11kx, we
follow the procedure of \citet{Dilday12} and fit multiple Gaussian
components to the \halpha\ and \hbeta\ emission lines. 
The \halpha\ profiles of our PTF11kx spectra are shown in
Figure~\ref{f:hal_zoom}. Clearly the profiles consist of a broader  
(FWHM $\approx 2000$~\kms) component combined with a narrow
(FWHM $< 200$~\kms), 
unresolved component. While P-Cygni profiles were seen in 
high-resolution spectra of PTF11kx \citep{Dilday12}, the spectra
presented herein have relatively low resolution, so we do not expect to
observe such subtle features.

\begin{figure*}
\centering$
\begin{array}{c}
\includegraphics[width=5in]{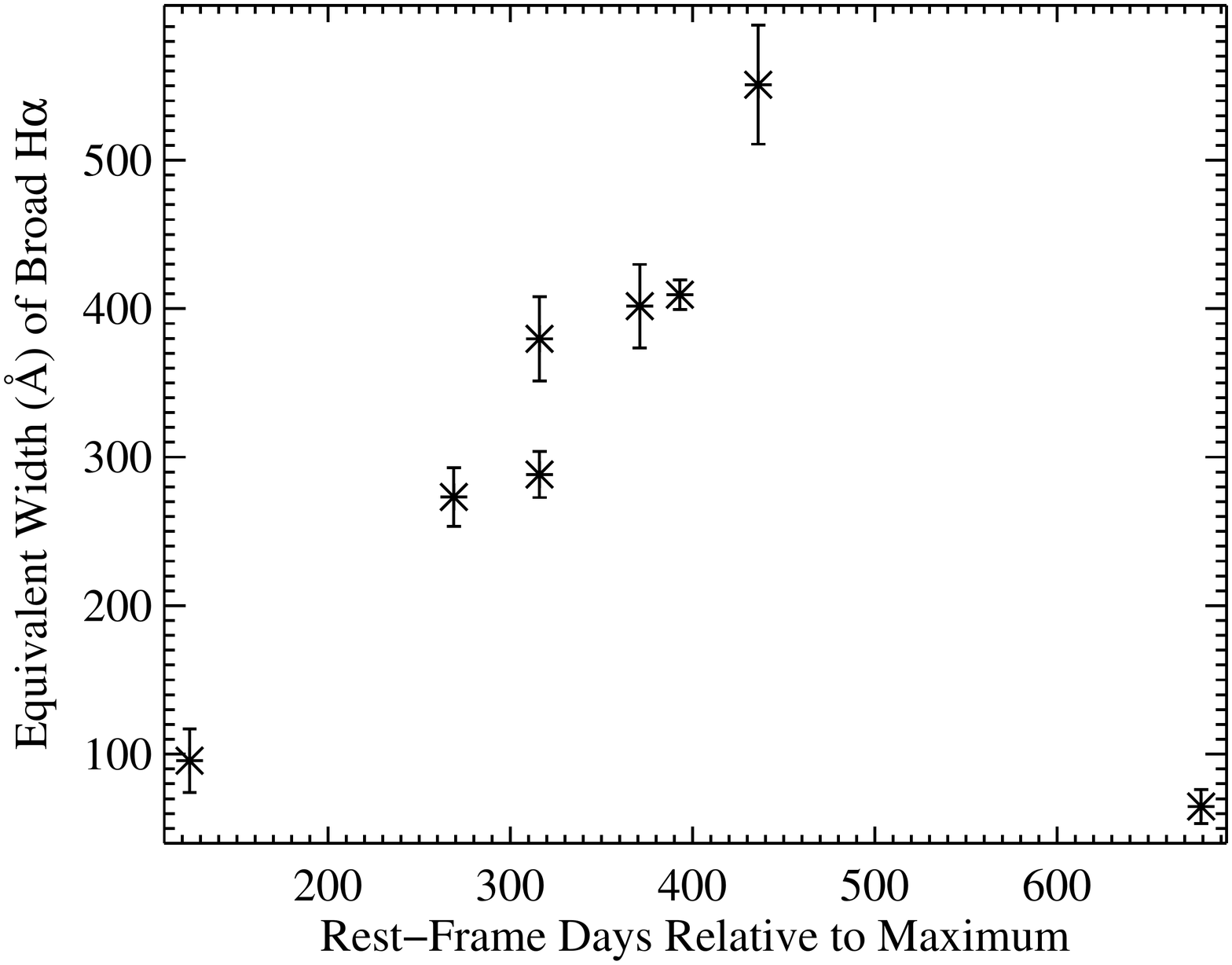} \\
\includegraphics[width=5in]{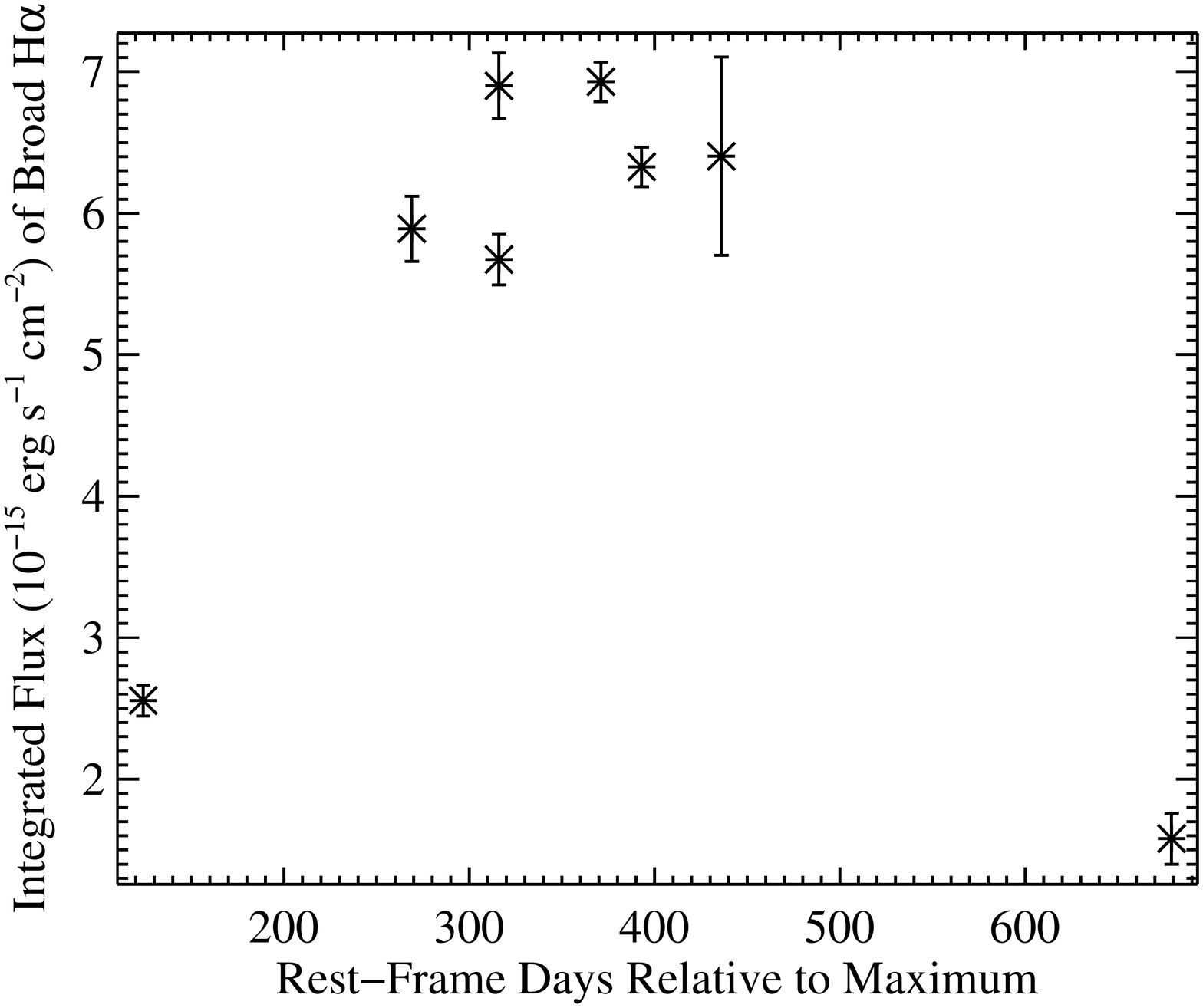} \\
\end{array}$
\caption{Temporal evolution of the EW ({\it top}) and flux ({\it bottom})
  of the broader \halpha\ 
  component.}\label{f:ew_b_hal}
\end{figure*}

Figure~\ref{f:ew_b_hal} shows the temporal evolution of the equivalent
width (EW; {\it top})
and flux ({\it bottom}) of the broader \halpha\ component. We find
that the broader \halpha\ component (after removing the narrow
component) is mostly symmetric and slightly blueshifted, perhaps due
to the CSM being accelerated somewhat by the SN ejecta. Its EW increases 
nearly linearly with time until our latest
spectrum (from \about680~d past maximum brightness), when the
EW drops significantly. The \halpha\ flux also generally increases 
with time until a large drop is measured in the last spectrum. Perhaps
the decrease in the strength of \halpha\ 
emission at +680~d is indicative of the SN ejecta finally overtaking
the majority of the CSM. In their $n = 8$ power-law model of SNe~IIn
(with $\dot{M} v^{-1} = 5\times10^{-4}$~\msun~yr$^{-1}$~km$^{-1}$~s),
the \halpha\ flux is predicted by \citet{Chevalier94} to decrease by a 
factor of \about2 from 1 to 2~yr after maximum brightness. The factor
of \about4 decrease shown by PTF11kx is larger than this prediction,
but it occurs at similar epochs. This may indicate that the density
structure of the CSM and/or mass-loss history of the PTF11kx system is
different than what was nominally assumed by \citet{Chevalier94}.

The \halpha\ profile in the spectrum from 680~d past maximum
exhibits a possible decrease in flux in the red wing as compared to the
blue wing. The dashed line on the bottom spectrum in
Figure~\ref{f:hal_zoom} is the reflection of the blue half of the
\halpha\ profile across the peak flux. This suppression of the red wing
has been seen in many SNe~IIn and is often interpreted as new dust
forming in the post-shock material \citep[e.g.,][]{Fox11,Smith12}. This
is also observed in all other SNe~Ia-CSM, but beginning much
earlier \citep[\about75--100~d past maximum brightness;][]{Silverman13:Ia-CSM}.

\begin{figure*}
\centering$
\begin{array}{c}
\includegraphics[width=5in]{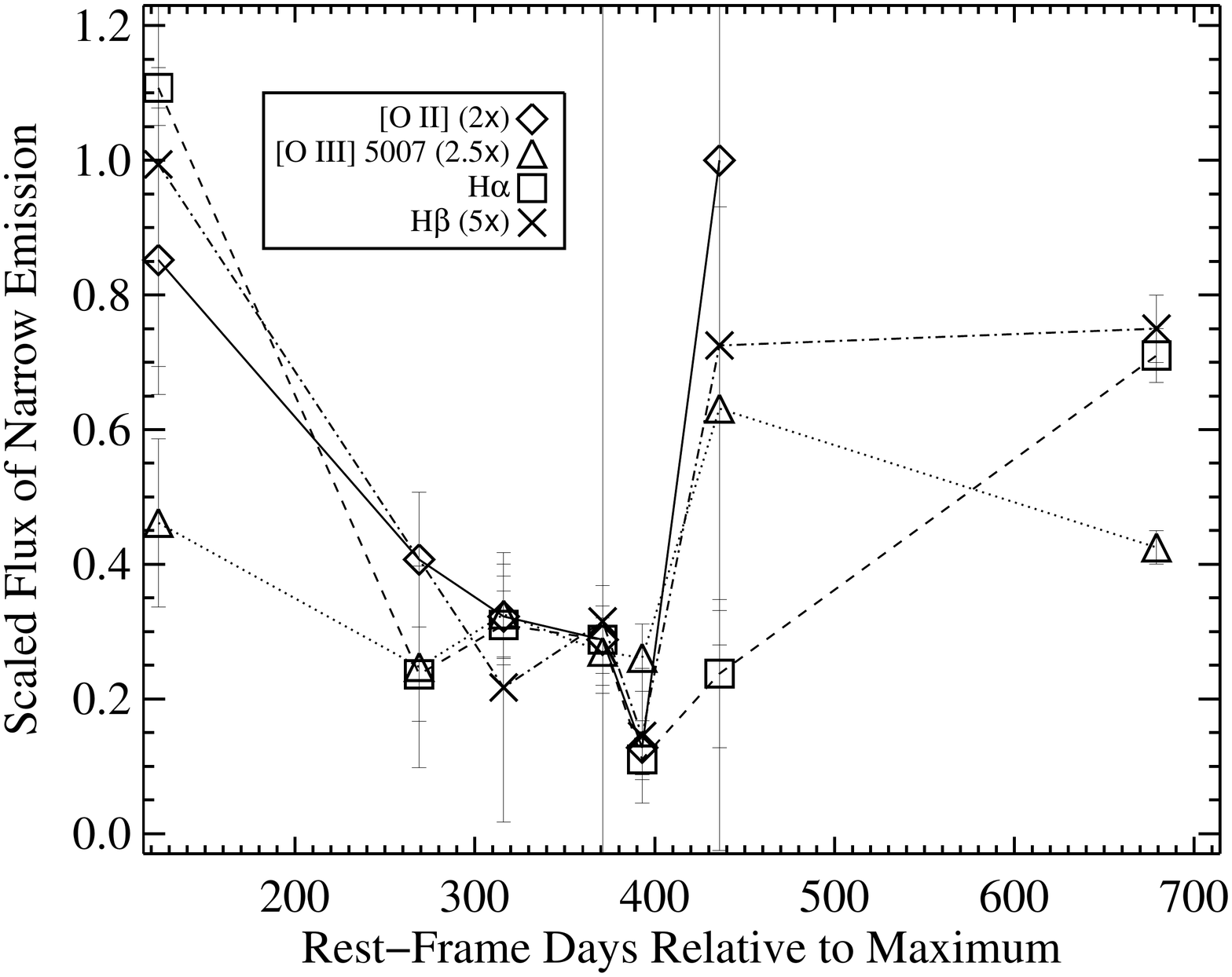} \\
\includegraphics[width=5in]{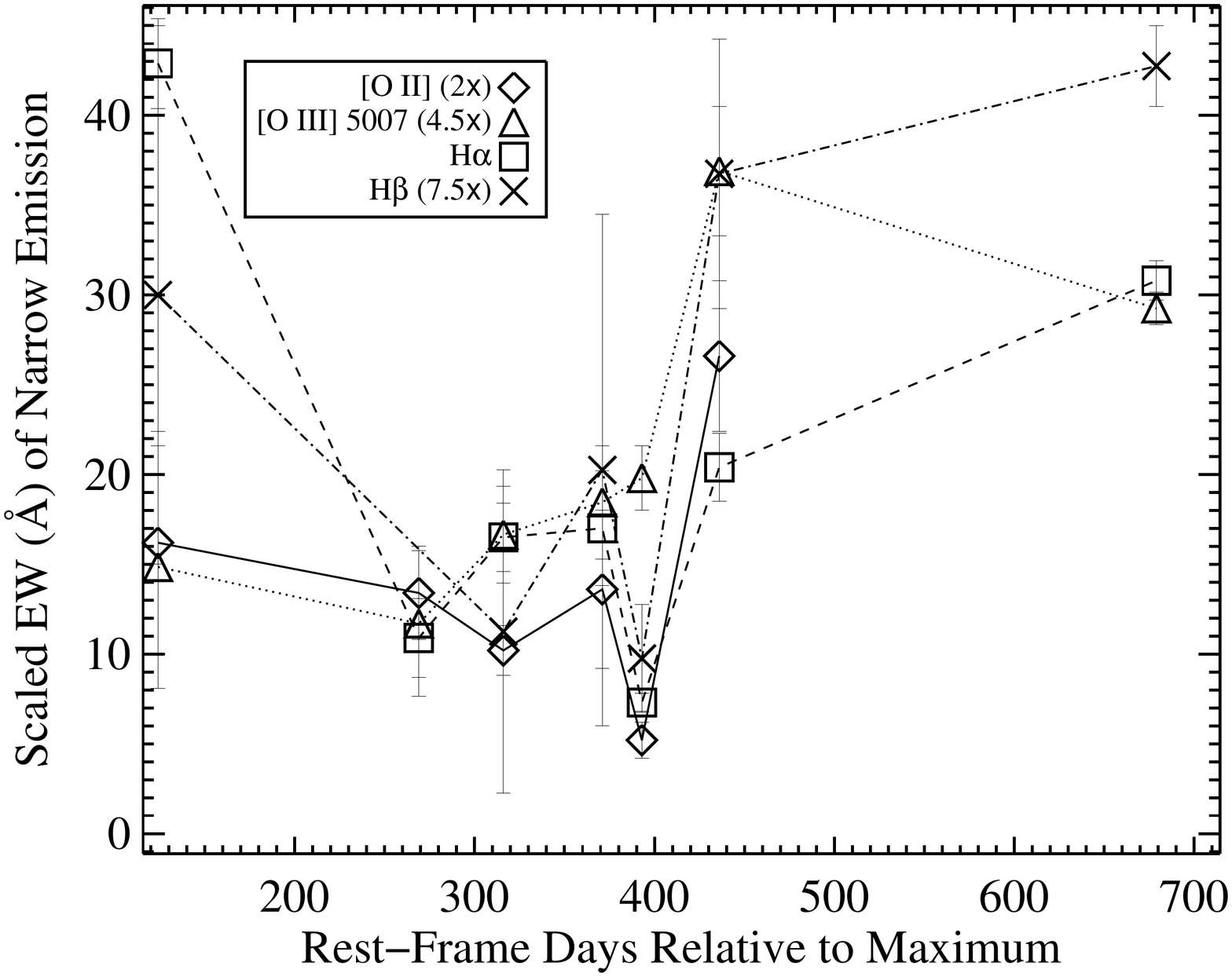} \\
\end{array}$
\caption{The temporal evolution of the flux ({\it top}) and EW ({\it
    bottom}) of the narrow 
      components of [\ion{O}{2}] $\lambda$3727 (diamonds and solid
      line), [\ion{O}{3}] $\lambda$5007 (triangles and dotted line),
      \halpha\ (squares and dashed line), and \hbeta\ (Xs and
      dot-dashed line). The values have been scaled to emphasize their 
      common evolution.}\label{f:ew_comp} 
\end{figure*}

Narrow emission lines of [\ion{O}{2}] $\lambda$3727, [\ion{O}{3}]
$\lambda$5007, and \hbeta\ are apparent in many of the spectra
shown in Figure~\ref{f:ptf11kx}. The temporal evolution of the flux of
these lines, along with the flux of the narrow component of \halpha\
emission, is displayed in the top panel of Figure~\ref{f:ew_comp}. The
flux values have been scaled (by factors of 2, 2.5, and 5 for
[\ion{O}{2}] $\lambda$3727, [\ion{O}{3}] $\lambda$5007, and \hbeta,
respectively) in order to emphasize their common evolution. The
bottom panel of Figure~\ref{f:ew_comp} shows the EWs of these narrow
features, again scaled (by factors of 2, 4.5, and 7.5 for [\ion{O}{2}]
$\lambda$3727, [\ion{O}{3}] $\lambda$5007, and \hbeta, respectively)
to highlight their common changes with time.

There may be some concern that slit losses, 
seeing effects, or contamination from nearby \ion{H}{2} regions
cause the observed variation in these lines with time. However, all
observations later than 300~d past maximum brightness were taken under
relatively good seeing conditions ($\la 1.1$\arcsec) using a 1\arcsec\
slit. In addition, the observations from 316 and 371~d past maximum
use position angles that differed by \about90\arcdeg, and yet they
yield nearly {\it identical} flux (and EW) measurements for all 4
narrow emission lines. On the other hand, the spectra from 371 and
393~d past maximum were obtained using the same position angle, but
these data yield significantly {\it different} flux (and EW) values
for each spectral feature. The flux measurements were obtained from
the spectra displayed in Figure~\ref{f:ptf11kx} after scaling each one
to nearly contemporaneous photometry of PTF11kx (or a fit to the linear
decline of the light curve) shown in Figure~\ref{f:lc} below.

The flux and EW of all four emission lines appear in
general to decrease, remain constant for \about150~d, then perhaps
increase again. However \halpha, which is a recominbation line and can
have a different timescale than the forbidden lines, sometimes
behaves significantly differently when compared with the other lines. This
could mean that the SN ejecta of PTF11kx are interacting with CSM that 
is clumpy or perhaps composed of multiple thin shells, which is
consistent with what was seen in the early-time spectra
\citep{Dilday12}. It seems unlikely to observe this behavior if the
ejecta were expanding into CSM with a monotonically increasing or
decreasing density profile, as one might expect from a with a
constant mass-loss rate. This conclusion is supported by the extreme
late-time overluminosity of PTF11kx (see
\S\ref{s:comparisons}). Nevertheless, we caution that perhaps some of this
variability could be caused by slit losses, seeing effects, or
contamination from nearby \ion{H}{2} regions.

When comparing both the broader and narrower components, \halpha\ is
significantly stronger than \hbeta\ at all epochs. The \halpha/\hbeta\
intensity ratio is $> 7$ in all of our late-time spectra of PTF11kx.
It varies throughout our observations, but it appears to generally
increase with time, peaking at a value of \about15 at $t \approx
680$~d. A large, time-variable \halpha/\hbeta\ intensity ratio seems
to be a hallmark of the SN~Ia-CSM class \citep{Silverman13:Ia-CSM}.

\begin{figure*}
\centering
\includegraphics[width=6in]{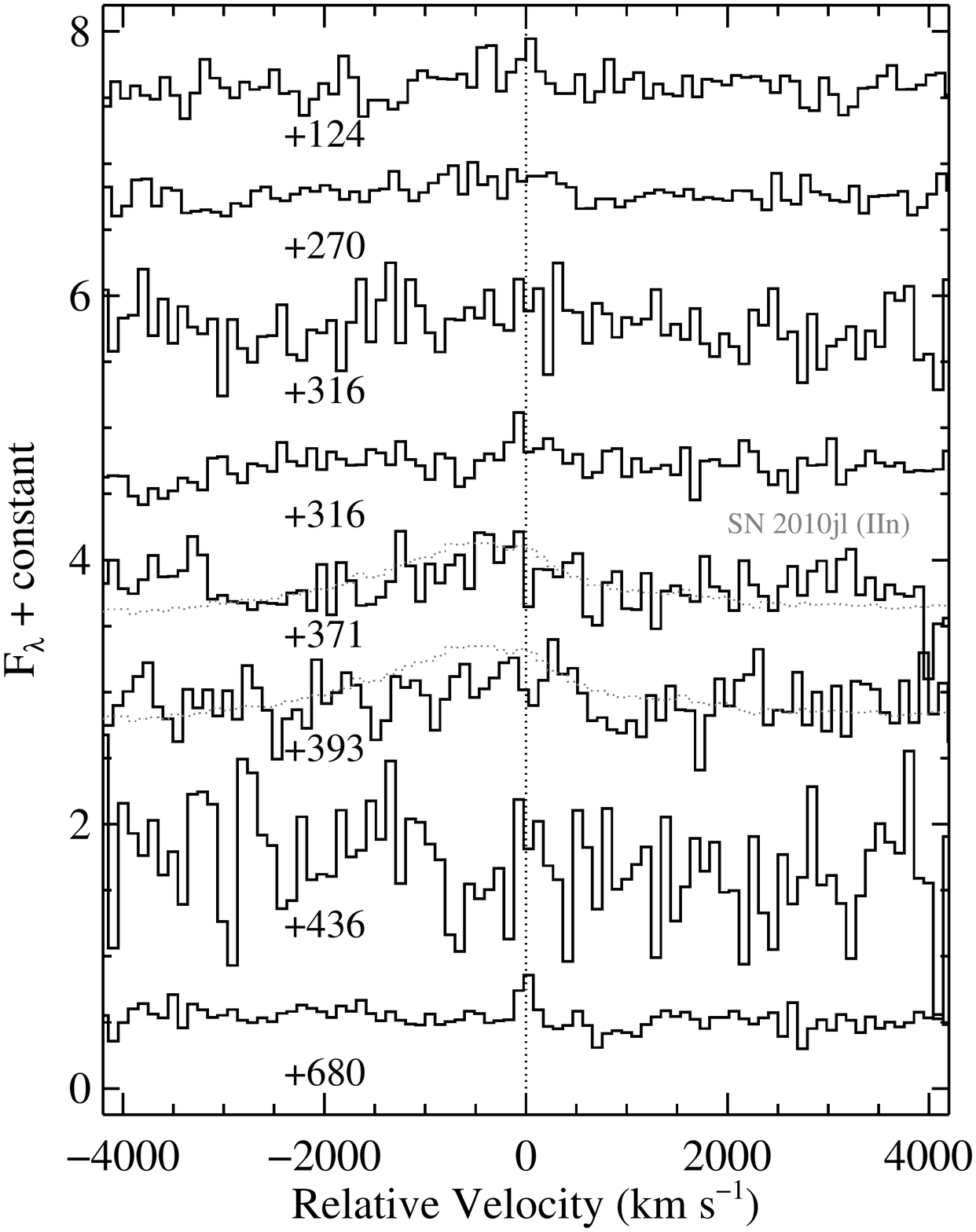}
\caption{The region near \ion{He}{1} $\lambda$5876 in our
  spectra of PTF11kx, labeled with age relative to maximum 
  brightness. Overplotted (gray dotted line) is the \ion{He}{1}
  $\lambda$5876 line of SN~2010jl at \about400~d after discovery
  \citep{Smith11}. The data have had their host-galaxy recession velocity
  removed and have been corrected for Galactic reddening. The dotted
  vertical line is the systemic velocity of
  PTF11kx.}\label{f:old_spec_he}
\end{figure*}

Another likely tell-tale sign of an object being a SN~Ia-CSM is weak
\ion{He}{1} emission \citep{Silverman13:Ia-CSM}. PTF11kx exhibited weak
\ion{He}{1} $\lambda$5876 at early times \citep{Dilday12}.
Figure~\ref{f:old_spec_he} shows the \ion{He}{1} 
$\lambda$5876 region in our late-time PTF11kx spectra; we see
hints of emission from this feature in some (but not all) cases. 
The clearest \ion{He}{1} $\lambda$5876 emission is
in the form of an unresolved narrow line in our final
spectrum at 680~d past maximum brightness. There is no evidence of emission
from \ion{He}{1} $\lambda$7065 in any of our spectra, except
possibly very weak, unresolved emission again in the 680~d spectrum. 

Overplotted (gray dotted line) on the data with the ``strongest'' resolved 
\ion{He}{1} emission in Figure~\ref{f:old_spec_he} is the 
\ion{He}{1} $\lambda$5876 profile at \about400~d after
discovery of SN~2010jl, a somewhat luminous SN~IIn that probably came
from a massive star \citep[thus likely {\it not} a SN~Ia of any
flavor;][]{Stoll11,Smith11}. The \ion{He}{1} $\lambda$5876 emission is
stronger and at higher S/N in SN~2010jl than any of the possible
\ion{He}{1} $\lambda$5876 detections in PTF11kx at similar epochs. The
most similar feature in the PTF11kx spectra is the possible resolved
line on day +371 (see also Fig.~\ref{f:ptf11kx}).

\section{Comparisons with Other Supernovae}\label{s:comparisons}

\begin{figure*}
\centering
\includegraphics[width=6.8in]{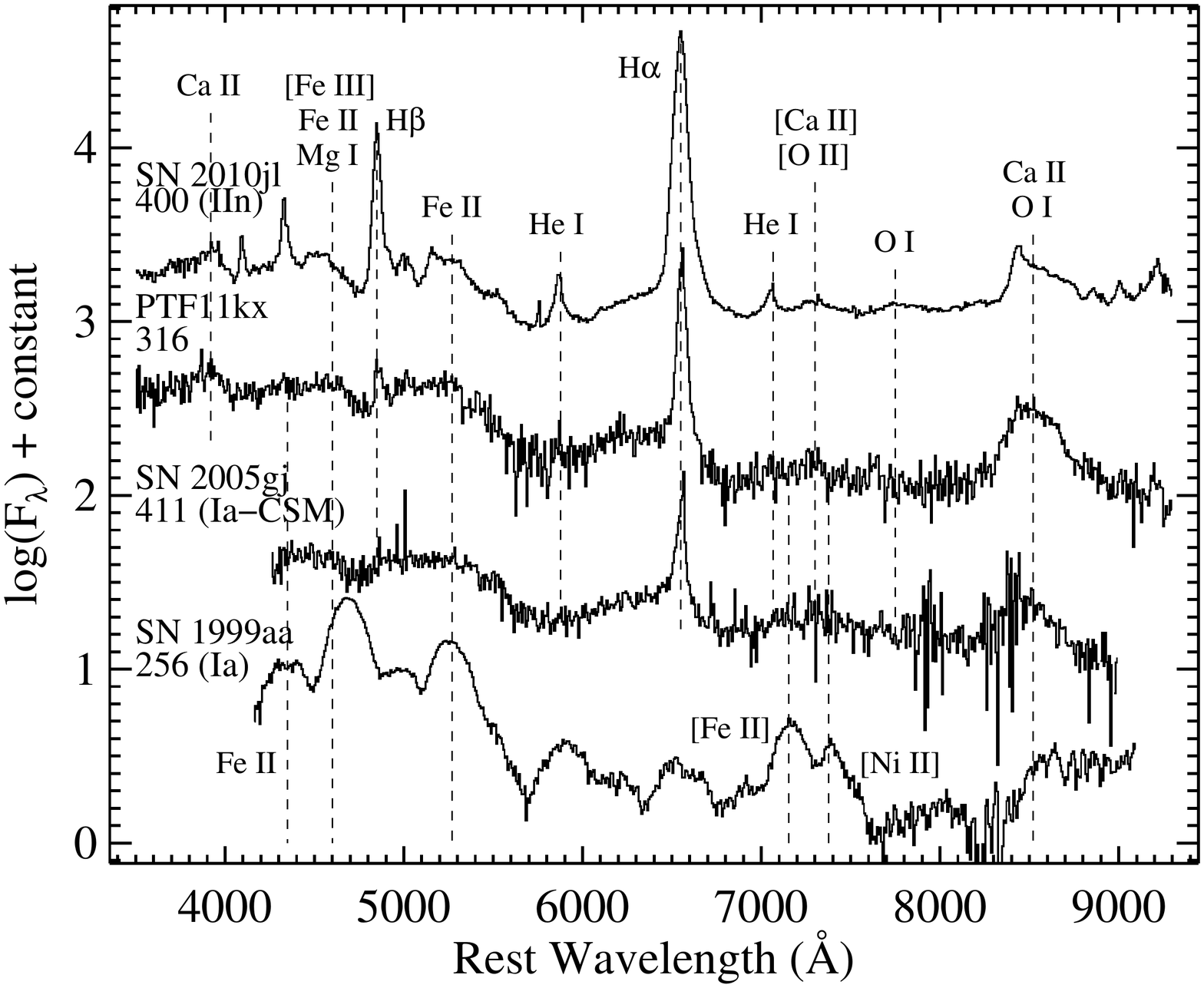}
\caption{The spectrum of PTF11kx from 316~d past maximum brightness,
  along with three other SNe: the luminous Type~IIn SN~2010jl from
  \about400~d after discovery \citep{Smith11}; one of the prototypical
  SNe~Ia-CSM, SN~2005gj, from 411~d after maximum brightness
  \citep{Silverman13:Ia-CSM}; and the somewhat overluminous Type~Ia
  SN~1999aa from 256~d 
  past maximum \citep{Silverman12:BSNIPI}. Major spectral features are
  labeled. The data have had their host-galaxy recession velocity
  removed and have been corrected for Galactic
  reddening.}\label{f:11kx_comp}  
\end{figure*}

In Figure~\ref{f:11kx_comp} we plot the spectrum of PTF11kx from 
316~d past maximum brightness, one of our highest S/N late-time
observations. In addition, we show late-time spectra of three other
SNe. 
At these late epochs, PTF11kx resembles the SNe~Ia-CSM more than any 
other SN type. Therefore, we follow the spectral feature identifications
that \citet{Deng04} used for the other prototypical SN~Ia-CSM, SN~2002ic 
(which are also valid for SN~2005gj).

The spectrum of PTF11kx is dominated by relatively narrow \halpha\
emission, which is of course also seen in SNe~2005gj and 2010jl (but
not in SN~1999aa). The \halpha\ emission is somewhat weaker in
PTF11kx relative to that of SN~2010jl, but comparable to that
of SN~2005gj. Similarly, narrow \hbeta\ emission is detected in
PTF11kx, but it is substantially weaker than \halpha\ (as mentioned in
\S\ref{s:analysis}). The strength of the \hbeta\ feature in PTF11kx
appears to be intermediate between those of SN~2010jl and
SN~2005gj. Higher-order Balmer emission is easily seen in SN~2010jl,
but completely undetected in PTF11kx and SN~2005gj.

The broad peak in PTF11kx near 3900~\AA\ and in PTF11kx and SN~2005gj
near 8500~\AA\ are almost certainly caused in part by \ion{Ca}{2}~H\&K
and the \ion{Ca}{2} near-IR triplet, respectively, and this latter
feature is one of the strongest in both spectra (with FWHM
\about10,000~\kms). Note that emission from these \ion{Ca}{2} 
complexes is also seen in SNe~IIn and SNe~Ia at late times. In
contrast to the +316~d spectrum of PTF11kx shown in
Figure~\ref{f:11kx_comp}, there is practically no emission from the
\ion{Ca}{2} near-IR triplet in the spectrum from 680~d past maximum
brightness (see Fig.~\ref{f:ptf11kx}). A model for SNe~Ia-CSM
that includes a cool dense shell made up of many small
fragments containing separate Fe-poor and Fe-rich zones predicts that 
\ion{Ca}{2} emission should disappear once the two zones become
fully mixed \citep{Chugai04}. This was seen to happen in SN~1999E at
between 139 and 361~d past maximum \citep{Rigon03}, and we find the
same behavior in PTF11kx but at a later epoch (between 371 and 680~d
past maximum).

In all of the spectra shown in Figure~\ref{f:11kx_comp}, the blue flux
dramatically increases shortward of \about5700~\AA. The largest increase in
flux at the blue end of the optical range appears in SN~1999aa, which
comes from nebular emission lines from very iron-rich ejecta, excited
by radioactivity. On the other hand, the smallest increase in blue
flux appears in SN~2010jl, where this blue ``quasi-continuum'' probably 
comes from numerous blended, narrow lines of IGEs excited by
the CSM interaction. PTF11kx (as mentioned in \S\ref{s:spectra}) and
SN~2005gj have intermediate amounts of increased blue flux, and since
we see evidence of strong CSM interaction in these objects, this
``quasi-continuum'' is most likely a stronger version of what is seen
in SN~2010jl and less like the nebular emission lines of more normal
SNe~Ia (such as SN~1999aa, also see below). As most of the optical
flux in PTF11kx shortward of \about5700~\AA\ comes from numerous IGE
multiplets, it is difficult to identify individual features in that
wavelength range (aside from \ion{Ca}{2}~H\&K and \hbeta), though some
relatively narrow \ion{Fe}{2} emission features can be tentatively
identified.

The nearly complete lack of emission at \about7700~\AA\ in
PTF11kx (as well as SN~2005gj) leads us to believe that there is
little to no oxygen present in their late-time spectra. Nebular oxygen
emission is a hallmark of core-collapse SNe, and it appears in the spectrum 
of SN~2010jl presented in Figure~\ref{f:11kx_comp}. Thus, this supports
the notion that PTF11kx, and all SNe~Ia-CSM in general, are in fact
genuine SNe~Ia. Furthermore, PTF11kx and SN~2005gj exhibit almost no
\ion{He}{1} (see also \S\ref{s:analysis}), while the spectrum of SN~2010jl
shows obvious emission lines from \ion{He}{1} $\lambda$5876 and
$\lambda$7065. The almost total lack of oxygen and helium emission in
the very late-time PTF11kx spectra imply that the broad, weak emission 
near 7300~\AA\ is caused either by [\ion{Ca}{2}], as it likely is in
SN~2010jl, or by [\ion{Fe}{2}] and [\ion{Ni}{2}], which is what is
seen in more typical SNe~Ia at these epochs. We favor the former option
since PTF11kx shows no convincing evidence for strong, broad emission
features due to [\ion{Fe}{3}], [\ion{Fe}{2}], or [\ion{Ni}{2}] in any
of the observations.

\begin{figure*}
\centering
\includegraphics[width=6.8in]{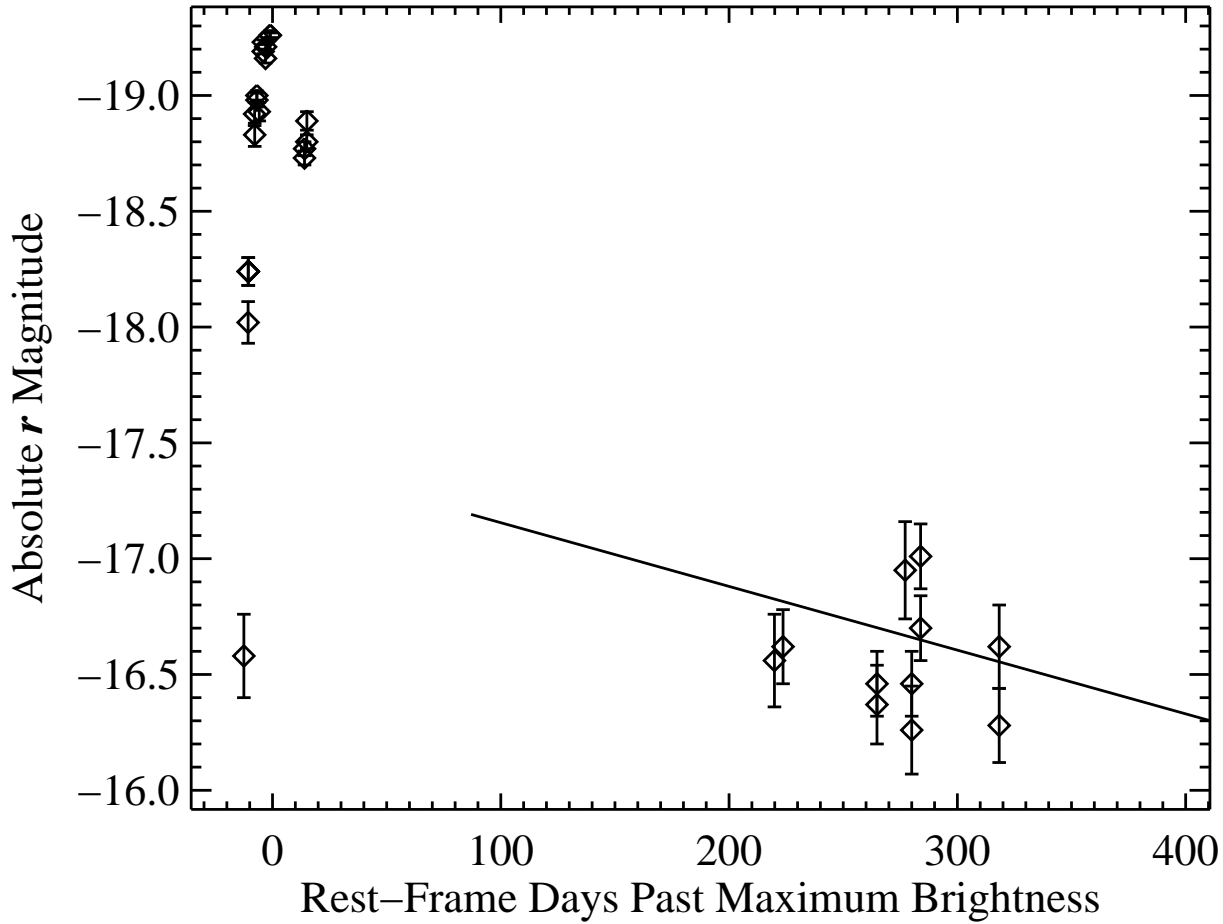}
\caption{The $r$-band light curve of PTF11kx with data near maximum
  brightness that come from \citet{Dilday12}. The solid line is
  the linear late-time decline of \about0.2~mag/(100~d).}\label{f:lc}
\end{figure*}

These forbidden emission features from IGEs that are typically seen in
more normal SNe~Ia at late times \citep[$t \ga 100$~d;
e.g.,][]{Silverman13:late} may not be detected in PTF11kx since they
have been diluted by broad-band continuum flux from the ongoing CSM
interaction. \citet{Dilday12} found that PTF11kx is \about16 times
more luminous than typical SNe~Ia at \about100~d past maximum
brightness, while more recent observations (shown in
Fig.~\ref{f:lc}) find that it is \about300 (\about2600) times more
luminous at \about460~d (\about680~d) past maximum. The extremely slow
decline rate at late times [\about0.2~mag/(100~d)] and the large
late-time luminosity of PTF11kx are inconsistent with a CSM model that
includes a wind with a constant mass-loss rate \citep[i.e., $\rho
\propto r^{-2}$,][]{Chugai04b}. This model was first proposed for
SN~2002ic by \citet{Chugai04b}, who also showed it to be inconsistent
with late-time observations of that object.

At these late epochs, PTF11kx is spectroscopically nearly identical to
SN~2005gj \citep[and to the other SNe~Ia-CSM as well;
e.g.,][]{Silverman13:Ia-CSM}. It looks quite unlike SN~1999aa, even  
though it closely resembled this object spectroscopically at early
times \citep{Dilday12}. 
Nebular spectra of other types of SNe~Ia were also compared with
those of PTF11kx, including the overluminous SN~1991T
\citep{Filippenko92:91T,Phillips92} and the normal SN~2011fe (Bianco
\etal in prep.), but as with SN~1999aa, they did not provide good
matches. A variety of other SNe~IIn were
compared to PTF11kx, and in all cases they either closely resembled
the spectrum of SN~2010jl presented in Figure~\ref{f:11kx_comp} or
looked completely different from both SN~2010jl {\it and}
PTF11kx. Finally, we compared PTF11kx to late-time spectra of the
normal Type~Ic SN~1994I \citep{Filippenko95} and the GRB-associated
Type~Ic SN~1998bw \citep{Patat01}; we found that while both have a
similar amounts of blue flux, the spectra of SNe~1994I and 1998bw are
dominated by [\ion{O}{1}] $\lambda$6300, [\ion{O}{2}] $\lambda7300$,
and \ion{Mg}{1}] $\lambda$4571 emission, which is in stark contrast to
the nebular spectra of PTF11kx \citep[and the other SNe~Ia-CSM;
e.g.,][]{Silverman13:Ia-CSM}.


\section{Conclusions}\label{s:conclusions}

PTF11kx was a SN~Ia with high levels of interaction with its CSM, and
it likely had a symbiotic nova progenitor \citep{Dilday12}. In this work
we have presented and analyzed late-time spectra of PTF11kx ranging
from 124 to 680~d past maximum brightness. While the SN shows very
little overall spectral evolution during these epochs (except for the
last spectrum), we find that the broader (\about2000~\kms) \halpha\
emission appears to increase in strength with time for \about1~yr,
after which time it decreases significantly, perhaps indicating that the
SN ejecta have overtaken most of the CSM. There are also indications
from the narrow \halpha\ component that the ejecta of PTF11kx 
are interacting with CSM that is clumpy or perhaps made up of multiple 
thin shells. This is consistent with what was seen in the early-time
spectra and the model of a symbiotic nova progenitor which had
multiple eruptions prior to the SN \citep{Dilday12}. We also find that
PTF11kx has an unusually large \halpha/\hbeta\ ratio that varies with
time, as well as extremely weak \ion{He}{1} emission (significantly
weaker than more normal SNe~IIn). Both of these are hallmarks of
SNe~Ia-CSM \citep{Silverman13:Ia-CSM}.

Aside from \halpha, the late-time spectra of PTF11kx also show strong,
broad \ion{Ca}{2} emission features (FWHM \about10,000~\kms). The
\ion{Ca}{2} emission disappeared nearly completely in the +680~d
spectrum, which agrees with a model of another SN~Ia-CSM
\citep{Chugai04}. This spectrum also indicates the possibility of
newly formed dust in the post-shock material as evidenced by a slight
decrease in the red wing of \halpha\ compared with the blue wing. At
all epochs little to no emission is detected from oxygen, \ion{He}{1},
and Balmer lines (aside from \halpha), which leads to a large observed
\halpha/\hbeta\ intensity ratio in PTF11kx. 

When comparing our late-time spectra of PTF11kx with those of other SNe, 
we have shown that it does not resemble SNe~Ic and only bears a passing
resemblance to more typical SNe~Ia (ones that follow the
\citealt{Phillips93} relation). PTF11kx shows a blue 
``quasi-continuum'' (due to numerous blended, relatively narrow lines
of \ion{Fe}{2}), but broad emission features from [\ion{Fe}{3}],
[\ion{Fe}{2}], and [\ion{Ni}{2}] (which are seen in more normal SNe~Ia
at late times) are absent. These features are completely diluted by
the broad-band continuum flux from the ongoing CSM interaction. The
extremely slow decline rate at late times, \about0.2~mag/(100~d), and
the large late-time luminosity, 4--9~mag brighter than a typical
SN~Ia, are almost certainly caused by the interaction between the SN
ejecta of PTF11kx and the CSM.

PTF11kx was a highly special case of a rare type of SN
(\about1\% of the SN~Ia population) being discovered
relatively early \citep[\about7~d after explosion;][]{Dilday12}. The
main observational properties of PTF11kx described above are shared by
other SNe~Ia-CSM, as well as by more typical SNe~IIn. Separating
SNe~Ia-CSM from SNe~IIn and studying the SN~Ia-CSM class in greater
detail are undertaken by \citet{Silverman13:Ia-CSM}. Despite this work
and the fortuitous discovery and extensive follow-up observations of
PTF11kx, questions still remain. Models of SNe~IIn, such
as those by \citet{Chevalier94}, seem to be somewhat applicable to
SNe~Ia-CSM, and models of other SNe~Ia-CSM 
\citep[e.g.,][]{Chugai04,Chugai04b} appear to match the observations
of PTF11kx. These models are a great starting point for future
theoretical work that we hope will utilize the observations and
analysis presented herein.

\begin{acknowledgments}
We would like to thank J. S. Bloom, K. Clubb, A. A. Miller, and A. Morgan 
for their assistance with some of the observations, B. Dilday, O. Fox,
and L. Wang for helpful discussions, and the anonymous referee for
providing comments and suggestions that improved the manuscript. 
We are grateful to the 
staffs at the WHT and the Keck Observatory for their support.
The WHT is operated on the island of La Palma by the Isaac Newton
Group in the Spanish Observatorio del Roque de los Muchachos of the
Instituto de Astrof\'{i}sica de Canarias. Some of the data presented
herein were obtained at the W. M. Keck Observatory, which is operated
as a scientific partnership among the California Institute of
Technology, the University of California, and the National Aeronautics
and Space Administration (NASA); the observatory was made possible by
the generous financial support of the W. M. Keck Foundation. The
authors wish to recognize and acknowledge the very significant
cultural role and reverence that the summit of Mauna Kea has always
had within the indigenous Hawaiian community; we are most fortunate to
have the opportunity to conduct observations from this mountain.
This research has made use of the NASA/IPAC Extragalactic Database
(NED) which is operated by the Jet Propulsion Laboratory, California
Institute of Technology, under contract with NASA. 
Funding for SDSS-III has been provided by the Alfred P. Sloan
Foundation, the Participating Institutions, the National Science
Foundation (NSF), and the U.S. Department of Energy Office of Science. 
The SDSS-III web site is http://www.sdss3.org/ . 
Supernova research by A.V.F.'s group
at U.C. Berkeley is supported by Gary and Cynthia Bengier, the Richard
and Rhoda Goldman Fund, the Christopher R. Redlich Fund, the TABASGO
Foundation, and NSF grants AST-0908886 and AST-1211916. Work
by A.G.-Y. and his group is supported by grants from the ISF, BSF, GIF,
and Minerva, an FP7/ERC grant, and the Helen and Martin Kimmel Award for
Innovative Investigation. M.S. acknowledges support from the Royal Society.
\end{acknowledgments}


\end{document}